\documentclass[aps,prl,twocolumn,superscriptaddress]{revtex4}
\usepackage{graphicx}
\usepackage{amssymb}
\usepackage{amsmath}
\usepackage{graphicx}
\usepackage{epsfig}
\usepackage{array}
\usepackage{color}
\usepackage{multirow}
\usepackage{ulem}

\begin{document}

\title{Inter-layer valence bonds and two-component theory for high-$T_c$
superconductivity of La$_{3}$Ni$_{2}$O$_{7}$ under pressure}
\author{Yi-feng Yang}
\email[]{yifeng@iphy.ac.cn}
\affiliation{Beijing National Laboratory for Condensed Matter Physics and Institute of
Physics, Chinese Academy of Sciences, Beijing 100190, China}
\affiliation{School of Physical Sciences, University of Chinese Academy of Sciences,
Beijing 100190, China}
\affiliation{Songshan Lake Materials Laboratory, Dongguan, Guangdong 523808, China}
\author{Guang-Ming Zhang}
\email[]{gmzhang@tsinghua.edu.cn}
\affiliation{State Key Laboratory of Low-Dimensional Quantum Physics and Department of
Physics, Tsinghua University, Beijing 100084, China}
\affiliation{Frontier Science Center for Quantum Information, Beijing 100084, China}
\author{Fu-Chun Zhang}
\email[]{fuchun@ucas.ac.cn}
\affiliation{Kavli Institute for Theoretical Sciences and CAS Center for Topological
Quantum Computation, University of Chinese Academy of Sciences, Beijing
100190, China }
\date{\today}

\begin{abstract}
The recent discovery of high-$T_{c}$ superconductivity in bilayer
nickelate La$_{3}$Ni$_{2}$O$_{7}$ under high pressure has stimulated great
interest concerning its pairing mechanism. We argue that the weak coupling
model from the almost fully-filled $d_{z^{2}}$ bonding band cannot give rise
to its high $T_{c}$, and thus propose a strong coupling model based on local inter-layer spin singlets of Ni-$d_{z^{2}}$ electrons due to their
strong on-site Coulomb repulsion. This leads to a minimal effective model
that contains local pairing of $d_{z^{2}}$ electrons and a considerable
hybridization with near quarter-filled itinerant $d_{x^{2}-y^{2}}$ electrons
on nearest-neighbor sites. Their strong coupling provides a
unique two-component scenario to achieve high-$T_{c}$ superconductivity. Our theory
highlights the importance of the bilayer structure of superconducting La$_{3}$Ni$_{2}$O$%
_{7}$ and points out a potential route for the exploration of
more high-$T_{c}$ superconductors.
\end{abstract}

\maketitle

\textbf{Introduction.}-- Exploration of high-$T_{c}$ superconductivity has
lasted for almost four decades since the first discovery of cuprate
superconductors \cite{Anderson-1987,Lee-Nagaosa-Wen-2006,Keimer-Zaanen-2015}%
. One of the main ideas is to start from a Mott insulator, suppress the
long-range antiferromagnetic order by doping, and create spin singlet pairs
whose phase coherence may eventually lead to the high-$T_{c}$
superconductivity \cite{Zhang-Rice}. However, attempts to replicate this
doped-Mott-insulator mechanism in nickelates have not been successful
despite of intensive experimental investigations \cite%
{Hayward-1999,Boris-2011,Disa-2015}. An advance is the discovery of the
infinite-layer (Nd,Pr)$_{1-x}$Sr$_{x}$NiO$_{2}$ nickelate superconductors
\cite{Li2019Nature,Li2020PRL,Zeng2020PRL,Osada-2020}, which have the desired $%
d^{9}$ configuration with almost half-filled Ni-$d_{x^{2}-y^{2}}$ orbitals
\cite{Anisimov-1999,Pickett-2004,Jiang-2020,Zhang2020PRB,Yang2022FP}. But superconductivity only appears in
thin films with the transition temperature below 40 K even under high
pressure \cite{Jinguang-2022}.

More recently, the bilayer La$_3$Ni$_2$O$_7$ bulk superconductor with a much
higher $T_c$ of about 80 K has been found under high pressure \cite%
{Sun2023,Yuan2023,Cheng2023,Zhang2023a}. Density functional theory (DFT)
calculations predicted a $d^{7.5}$ configuration with an almost fully-filled
$d_{z^2}$ bonding band and two $d_{x^2-y^2}$ bands near
quarter-filling \cite{Sun2023,Pickett-2011}, which is far from the Mott
regime. Although many theoretical works \cite%
{Luo2023,Dagotto2023,Qianghua2023,Kuroki2023,Jiangping2023,Shen2023,
WeiWu,Werner2023,Cao2023arXiv,FanYang2023} have been proposed, the key question
concerning its high-$T_c$ pairing mechanism remains open.

It is known that the highest $T_c$ of cuprate superconductors is achieved in
the trilayer structure of CuO$_2$ planes within a unit cell \cite%
{Scott1994,Chakravarty2004,Iyo2007}. Experiments have shown that the outer
and inner CuO$_2$ planes have different hole concentrations \cite%
{Ideta2010,Ideta2021}. While the outer planes are heavily hole-doped, the
inner one is only slightly doped. A composite scenario had been put forward
by Kivelson to understand how this might lead to the highest $T_c$ \cite%
{Kivelson2002}. This scenario contains a pairing component with a
high pairing scale $\Delta$ but small or zero superfluid stiffness, and a
metallic component with no pairing but high phase stiffness. The two
components are strongly coupled by a tunneling matrix element or
hybridization. Under proper conditions, a high $T_c \sim \Delta/2$ can be
reached \cite{Berg2008PRB} by assuming the superconducting transition as the
Kosterlitz-Thouless phase transition \cite{Emery-Kivelson}. In a recent
experimental work, Luo \textsl{et al.} argued that the outer and inner
planes in trilayer Bi$_2$Sr$_2$Ca$_2$Cu$_3$O$_{10+\delta}$ (Bi2223) cuprate
superconductors play the role of the pairing and metallic components,
respectively, and the unusual phase diagram in which $T_c$ keeps nearly constant
in the overdoped region might thus be well explained \cite{Luo2022NP}.

Although the application of Kivelson's composite scenario to the cuprate superconductors may be debated, it motivated us to propose a more realistic two-component theory for the high-$T_c$ mechanism in superconducting La$_3$Ni$_2$O$_7$
under high pressure. We argue in this work that the weak coupling picture starting from
the DFT single particle band structures with the almost fully-filled $d_{z^2}$
bonding band may not be able to give the high $T_c$. Rather, the strong
on-site Coulomb repulsion favors almost half-filled and localized $d_{z^2}$
electrons on the Ni ions. This, together with the special bilayer structure
of La$_3$Ni$_2$O$_7$, creates an inter-layer superexchange interaction of $%
d_{z^2}$ electrons through the apical O-$p_z$ orbital, and further induces local
spin singlets with a large pairing energy. We show that their hybridization
with near quarter-filled itinerant $d_{x^2-y^2}$ electrons on nearest-neighbor sites may explain the high-$T_c$ superconductivity in pressurized La$_3$Ni$_2$O$_7$. A minimal effective model is then constructed for more
elaborated investigations. Our scenario differs from the picture of the
one-band Hubbard model or $t$-$J$ model for cuprate superconductors \cite%
{Anderson-1987,Lee-Nagaosa-Wen-2006,Keimer-Zaanen-2015} and points out an
alternative route to realize the high-$T_c$ superconductivity by doping into the spin-singlet array \cite{Gao-Lu-Xiang2015}.

\textbf{Electronic band structures.}-- We start with considering the basic
lattice structure of superconducting La$_3$Ni$_2$O$_7$ under high pressure,
which contains two layers of Ni-O octahedra with shared apical O. The Ni-$d$
orbitals are split into fully-filled $t_{2g}$ orbitals, which are irrelevant
for the low-energy physics, and partially-filled $e_g$ orbitals. DFT
calculations \cite{Sun2023,Pickett-2011} predict dominant Ni-$d_{z^2}$ and
Ni-$d_{x^2-y^2}$ characters near the Fermi energy. These two orbitals are
orthogonal on individual Ni ion, but hybridize strongly between the in-plane
nearest-neighbor sites through the Ni-O-Ni bond. Due to the bilayer
structure, the Ni-$d_{z^2} $ orbitals coupled through the shared apical O-$%
p_z$ orbital along $z$ axis are further split into bonding and anti-bonding
states \cite{Shen2023}.

\begin{figure}[t]
\begin{center}
\includegraphics[width=0.45\textwidth]{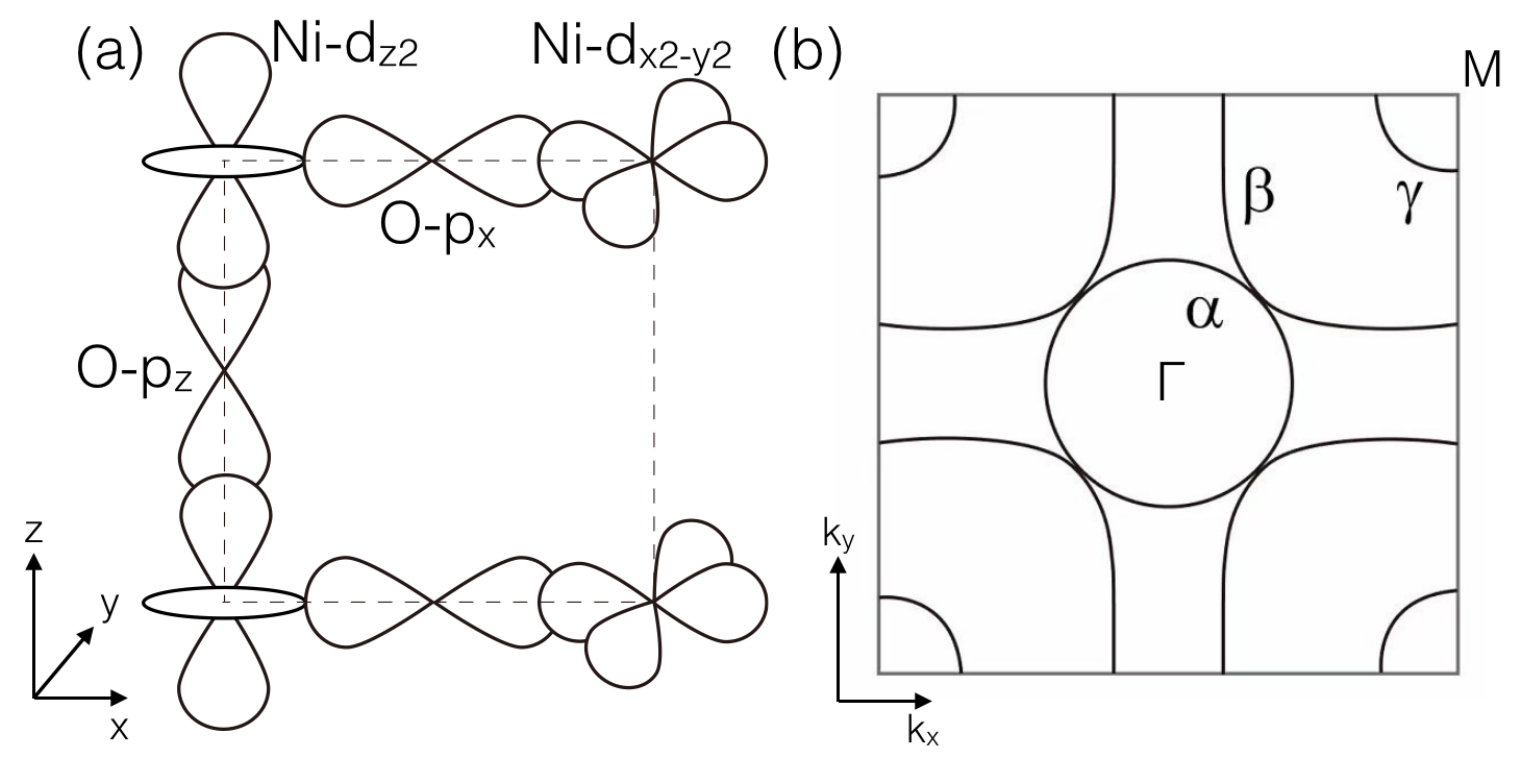}
\end{center}
\caption{(a) A schematic diagram of the wave function overlap of
bilayer La$_3$Ni$_2$O$_7$ under high pressure, showing the Ni-$%
d_{z^2}$ and $d_{x^2-y^2}$ orbitals and the O-$p$ orbitals. For clarity,
only one of the $d_{z^2}$ and $d_{x^2-y^2}$ orbitals is plotted on each Ni-ion. The $y$ direction is similar to
the $x$ direction and not shown. (b) A tentative plot of the DFT Fermi
surfaces, showing the $\protect\gamma$ hole pocket from the $d_{z^2}$
bonding band and the $\protect\alpha/\protect\beta$ Fermi surfaces from the
hybridized $d_{x^2-y^2}$ bands.}
\label{fig1}
\end{figure}

As illustrated in Fig. \ref{fig1}(a), we first construct the noninteracting
Hamiltonian of Ni-$d_{z^2}$ and $d_{x^2-y^2}$ orbitals for the bilayer
structure after integrating out the O degrees of freedom:
\begin{equation}
\begin{split}
H_0&=-\sum_{l(ij)
s}t^d_{ij}d^\dagger_{lis}d_{ljs}-t_\perp\sum_{is}(d^%
\dagger_{1is}d_{2is}+h.c.) \\
&-\sum_{l(ij)s}t^c_{ij}c^\dagger_{lis}c_{ljs}-\sum_{l\langle
ij\rangle s}(V_{ij}d^\dagger_{lis}c_{ljs}+h.c.),
\end{split}
\label{eqH0}
\end{equation}
where $d_{lis}$ ($c_{lis}$) annihilates a Ni-$d_{z^2}$ ($d_{x^2-y^2})$
electron of spin $s$ at site $i$ on layer $l$, $t_\perp$ accounts for the
inter-layer hopping of $d_{z^2}$ electrons via the shared apical O-$p_z$
orbital, $t^d_{ij}$ ($t^c_{ij}$) gives the onsite energy and in-plane hopping integral of $%
d_{z^2} $ ($d_{x^2-y^2})$, and $V_{ij}$ describes their hybridization
between nearest-neighbor sites which is positive along $x$ direction and negative along $y$
direction. For simplicity, we have neglected all other
inter-layer hopping terms that are supposed to be small. In momentum
space, the Hamiltonian may be rewritten as
\begin{equation}
H_0=\sum_{\mathbf{k}s}\Psi _{\mathbf{k}s}^{\dagger }\left(
\begin{array}{cccc}
\epsilon^c_\mathbf{k} & 0 & -V^*_\mathbf{k} & 0 \\
0 & \epsilon^c_\mathbf{k} & 0 & -V^*_\mathbf{k} \\
-V_\mathbf{k} & 0 & \epsilon^d_\mathbf{k} & -t_\perp \\
0 & -V_\mathbf{k} & -t_\perp & \epsilon^d_\mathbf{k}%
\end{array}%
\right) \Psi _{\mathbf{k}s},
\end{equation}
where $\Psi^\dagger_{\mathbf{k}s}=(c^\dagger_{1\mathbf{k}s}, c^\dagger_{2%
\mathbf{k}s}, d^\dagger_{1\mathbf{k}s}, d^\dagger_{2\mathbf{k}s})$, $%
\epsilon^d_\mathbf{k}$ and $\epsilon^c_\mathbf{k}$ are the dispersions of $%
d_{z^2}$ and $d_{x^2-y^2}$ electrons, respectively, and $V_\mathbf{k}%
=V(\cos\mathbf{k}_x-\cos\mathbf{k}_y)$ is their momentum-dependent
hybridization. The inter-layer coupling $t_\perp$ is site independent along
the planar directions and therefore remains a constant.

The above Hamiltonian can be easily diagonalized and gives an almost
fully-occupied $d_{z^2}$ bonding band and two $d_{x^2-y^2}$ bands around the
Fermi energy. As predicted by DFT, the $d_{z^2}$ antibonding band is above
the Fermi energy. The resulting Fermi surfaces in
the two-dimensional Brillouin zone are given in Fig. \ref{fig1}(b), where
the $d_{z^2}$ bonding band contributes a small hole pocket $\gamma$ around
M, while the two $d_{x^2-y^2}$ bands hybridize with the $d_{z^2}$ bands and
contribute the $\alpha$ and $\beta$ Fermi surfaces.

\textbf{Weak coupling picture.}-- The weak coupling picture starts from the
above DFT Fermi surfaces. Hence, the $d_{z^2}$ antibonding band is
irrelevant and may be in principle projected out to give an effective
three-band model:
\begin{equation}
H^{\text{3band}}_0=\sum_{\mathbf{k}s}\Phi _{\mathbf{k}s}^{\dagger }\left(
\begin{array}{ccc}
\epsilon^c_\mathbf{k} & 0 & -V^*_\mathbf{k}/\sqrt{2} \\
0 & \epsilon^c_\mathbf{k} & -V^*_\mathbf{k}/\sqrt{2} \\
-V_\mathbf{k}/\sqrt{2} & -V_\mathbf{k}/\sqrt{2} & \epsilon^d_\mathbf{k}%
-t_\perp%
\end{array}%
\right) \Phi _{\mathbf{k}s},
\end{equation}
where $\Phi_{\mathbf{k}s}=(c_{1\mathbf{k}s}, c_{2\mathbf{k}s}, d_{+\mathbf{k}%
s})^{T}$, and $d_{\pm\mathbf{k}s}=(d_{1\mathbf{k}s}\pm d_{2\mathbf{k}s})/%
\sqrt{2}$ describe the bonding (+) and antibonding (-) bands, respectively.
To achieve high $T_c$ of about 80 K, the pairing interaction should be
sufficiently strong. However, because the $\gamma$ band is almost fully-filled and the $\alpha$/$\beta$ bands are far from the Mott regime, the
effective three-band model contains only weak electronic correlations. As
may be seen in Fig. \ref{fig1}(b), none of the Fermi surfaces is well nested to
support a strong pairing interaction for the high-$T_c$ superconductivity.

\textbf{Effect of electronic correlations.}-- To include the correlation
effect, we start from the original $H_{0}$ in Eq. (\ref{eqH0}) and construct
an interacting Hamiltonian:
\begin{equation}
H=H_{0}+U\sum_{li}n_{li\uparrow }^{d}n_{li\downarrow }^{d},  \label{eqHU}
\end{equation}%
where $U$ is the onsite Coulomb repulsion of $d_{z^{2}}$ electrons, and the
Coulomb repulsion for $d_{x^{2}-y^{2}}$ orbital has been ignored because of
their quarter filling. Then the final Hamiltonian $H$ describes two coupled
layers of the periodic Anderson lattice model. For more realistic
calculations, we may also include a local Hund's rule coupling between $%
d_{z^{2}}$ and $d_{x^{2}-y^{2}}$ orbitals: $-J_{H}\sum_{li}\mathbf{S}%
_{li}\cdot \mathbf{s}_{li}$, where the spin operators are defined as $%
\mathbf{S}_{li}=\frac{1}{2}\sum_{ss'}d_{lis }^{\dagger }%
\boldsymbol{\sigma }_{ss'}d_{lis' }$ and $\mathbf{s}_{li}=\frac{%
1}{2}\sum_{ss'}c_{lis }^{\dagger }\boldsymbol{\sigma }%
_{ss'}c_{lis'}$, with $\boldsymbol{\sigma }$ being the Pauli
matrices. It has been shown that the Hund's rule coupling tends to compete
with the hybridization and promote the quasiparticle flat bands \cite%
{Cao2023arXiv}.

To see the effect of $U$, we first consider a toy model of two coupled Ni-$%
d_{z^2}$ orbitals:
\begin{equation}
H_\text{toy}=-t_\perp\sum_{s}(d^\dagger_{1s}d_{2s}+h.c.)+U\sum_{l}n_{l%
\uparrow}n_{l\downarrow},
\end{equation}
For half filling, we may rewrite the above Hamiltonian in a matrix form in
the subspace of a total electron number $N_d=2$ and a total spin $S^z=0$:
\begin{equation}
H_\text{toy}=\left(
\begin{array}{cccc}
0 & 0 & -t_\perp & -t_\perp \\
0 & 0 & t_\perp & t_\perp \\
-t_\perp & t_\perp & U & 0 \\
-t_\perp & t_\perp & 0 & U%
\end{array}
\right).
\end{equation}
It may then be diagonalized to give the ground state energy $E_0=(U-\sqrt{%
U^2+16 t_\perp^2})/2$ and the wave function:
\begin{equation}
|\text{GS}\rangle=\frac{1}{\sqrt{1+\alpha^2}}\left(\frac{|20\rangle+|02%
\rangle}{\sqrt{2}} +\alpha\frac{|\uparrow\downarrow\rangle-|\downarrow%
\uparrow\rangle}{\sqrt{2}}\right) ,
\end{equation}
which is a superposition of the inter-layer spin singlet and two
doubly-occupied states. Their relative weight is determined by the parameter
$\alpha=\sqrt{x^2+1}+x$ with $x=U/4t_\perp$. Thus for $U\ll 4t_\perp$, we
have $\alpha\rightarrow 1$ and $|\text{GS}\rangle\approx (|\text{$\uparrow$}%
0\rangle+|0\text{$\uparrow$}\rangle)(|\text{$\downarrow$}0\rangle+|0\text{$%
\downarrow$}\rangle)/2$, which recovers the doubly-occupied $d_{z^2}$
bonding state, while the antibonding state is unoccupied with a higher
energy. For $U\gg 4t_\perp$, we have instead $\alpha^{-1}\rightarrow 0$ and $%
|\text{GS}\rangle\approx \frac{|\uparrow\downarrow\rangle-|\downarrow%
\uparrow\rangle}{\sqrt{2}}$, which is an inter-layer spin singlet. The
ground state energy of the spin singlet is $E_0\approx -4t_\perp^2/U$, which
corresponds to the superexchange energy. Taking for example $U=5$ eV and $%
t_\perp=0.6$ eV \cite{Luo2023}, the weight of the doubly-occupied state is
only about 5 percent. Therefore, even for such a large $t_\perp$ from DFT,
the on-site Coulomb repulsion can still effectively suppress double
occupancy, enforce almost fully localized Ni-$d_{z^2}$ electrons, and
produce an inter-layer spin singlet. It is therefore necessary to start from
localized $d_{z^2}$ electrons instead of the almost fully-filled $d_{z^2}$
bonding state.

\textbf{Strong coupling picture.}-- Our analysis suggests a strong coupling
picture with local inter-layer spin singlets of $d_{z^2}$
electrons. They are further hybridized with the near quarter-filled $d_{x^2-y^2}$
bands. The hybridization may cause some electron transfer 
from $d_{z^2}$ to $d_{x^2-y^2}$ orbitals, which is equivalent
to introduce self-doped holes in $d_{z^2}$ orbitals. On the other hand, despite of the strong hybridization, the $d_{z^2}$ spins cannot be fully screened by $d_{x^2-y^2}$ electrons because of their quarter filling. We are therefore left with both strongly renormalized hybridization bands and a large residual inter-layer superexchange coupling that
may mediate the electron pairing \cite{Cao2023arXiv}.

\begin{figure}[t]
\begin{center}
\includegraphics[width=0.4\textwidth]{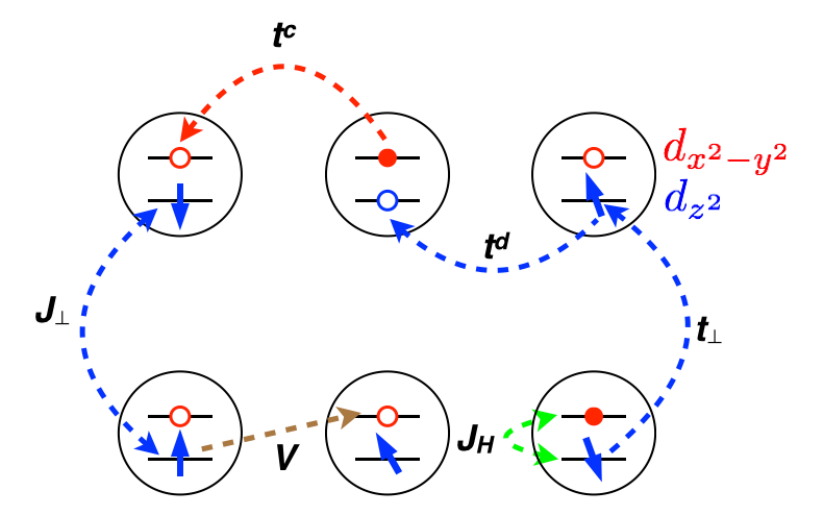}
\end{center}
\caption{Illustration of some main low-energy processes. $J_\perp$ ($t_\perp$) denotes the inter-layer superexchange interaction
(hopping) between Ni-$d_{z^2}$ electrons, $t^c$ and $t^d$ are the hopping parameters
of $d_{x^2-y^2}$ and $d_{z^2}$ electrons, respectively, $J_H$ is their local
Hund's rule coupling, and $V$ is their in-plane nearest-neighbor
hybridization.}
\label{fig2}
\end{figure}

Thus, a minimal effective model may be derived by projecting out only the
double occupancy of Ni-$d_{z^2}$ orbital in the interacting Hamiltonian (\ref%
{eqHU}):
\begin{equation}
\begin{split}
H_{\text{eff}}&=J_\perp\sum_i\mathbf{S}_{1i}\cdot\mathbf{S}%
_{2i}-t_\perp\sum_{is}(d^\dagger_{1is}d_{2is}+h.c.) \\
&-\sum_{l(ij)s}t^c_{ij}c^\dagger_{lis}c_{ljs}-V\sum_{lis}(d^\dagger_{lis}\tilde{c}_{lis}+h.c.),
\end{split}
\label{eqHJ}
\end{equation}
where $J_\perp$ is the inter-layer superexchange coupling of $d_{z^2}$
spins and we have defined $\tilde{c}_{lis}=\frac12(c_{l,i+\mathbf{x},s}+c_{l,i-\mathbf{x},s}-c_{l,i+%
\mathbf{y},s}-c_{l,i-\mathbf{y},s})$ to reflect the nearest-neighbor hybridization. Similar to the $t$-$J$ model for cuprate superconductors, the above
model is subject to a local constraint $n^{d}_{li}=\sum_{s}d^%
\dagger_{lis}d_{lis}=1-\delta_d \le1$, where $\delta_d$ counts the self-doped hole density on $d_{z^{2}}$ orbital. We have dropped the Hund's rule coupling, whose
effect is partly included in the renormalization of the hybridization parameter
\cite{Cao2023arXiv}. Other parameters may also be modified by
electronic correlations. The small $d_{z^2}$ intra-layer hopping is ignored in the minimal model for
simplicity. As in heavily hole-doped cuprates, the intra-layer superexchange
interaction is small and dropped because of the near quarter-filled $d_{x^2-y^2}$. Figure \ref{fig2} gives an illustration of these terms in the low-energy
physics. Simply from this effective model, we can already see that La$_3
$Ni$_2$O$_7$ is a unique system with special characters for exploring exotic
many-body quantum phenomena.

\textbf{Route to high-$T_c$ superconductivity.}-- Although somewhat
different, the similarity compared to the $t$-$J$ model for hole-doped
cuprates suggests a candidate route towards high-$T_c$ superconductivity,
with the inter-layer $d_{z^2}$ spin singlets providing a high pairing energy
scale given by the inter-layer superexchange interaction $J_\perp$. To see how this works, we first decouple the superexchange term in the Hamiltonian (\ref{eqHJ}):
\begin{equation}
J_{\perp }\sum_{i}\mathbf{S}_{1i}\cdot \mathbf{S}_{2i}\rightarrow -\Delta
_{d}^{\ast }\psi _{i}^{d}+h.c.,
\end{equation}%
where $\psi _{i}^{d}=\frac{1}{\sqrt{2}}\sum_{ss'}d_{1is}(i\sigma _{ss'}^{y})d_{2is'}$ is the $%
d_{z^{2}}$ local inter-layer spin singlet pair and $\Delta _{d}=\frac{3}{4}J_{\perp
}\langle \psi _{i}^{d}\rangle $ gives the mean-field self-consistent equation. 
A second-order perturbation with $V$ yields the (static) $d_{x^2-y^2}$ pairing term:
\begin{equation}
\begin{split}
H_{\Delta }&=-g_t\frac{\sqrt{2}V^2}{\Delta_d}\sum_{i}\left[\tilde{c}_{1is}(i\sigma^y_{ss'})\tilde{c}_{2is'}+h.c.\right]\\
&=-\sum_{\mathbf{k}}\left[ \Delta _{c}^{\ast }(%
\mathbf{k})\psi _{\mathbf{k}}^{c}+h.c.\right] ,
\end{split}
\end{equation}%
where $\psi _{\mathbf{k}}^{c}=\frac{1}{\sqrt{2}}\sum_{ss'}c_{1%
\mathbf{k}s }(i\sigma _{ss'}^{y})c_{2\mathbf{-k}s'}$ for
$d_{x^{2}-y^{2}}$ electrons, $\Delta _{c}(\mathbf{k})=\tilde{\Delta}%
_{c}(\cos \mathbf{k}_{x}-\cos \mathbf{k}_{y})^{2}$ with the proximity
induced pairing field $\tilde{\Delta}_{c}\sim g_{t}V^{2}/\Delta _{d}$, and  
$g_{t}\sim \delta_{d}$ accounts for Gutzwiller projection to exclude the $d_{z^2}$ double occupancy as in the $t$-$J$ model. This predicts nodeless $s$-wave pairing on the $d_{z^2}$ Fermi surface and extended $s$-wave pairing with possible nodes or gap minima along the zone diagonal on the $d_{x^{2}-y^{2}}$ Fermi surfaces. To simplify the discussions, we have ignored the renormalized inter-layer hopping term $t_{\perp }$, which determines the detailed (anti-)bonding and orbital properties of each Fermi surface but does not affect the primary pairing mechanism. In reality, $\alpha$ and $\gamma$ Fermi surfaces are from the bonding bands of $d_{x^2-y^2}$ or $d_{z^2}$ orbitals so that their gap functions have the same sign, while the $\beta$ Fermi surface is from the antibonding band of hybridized $d_{x^2-y^2}$ and $d_{z^2}$ orbitals and its gap function has an opposite sign \cite{Qin2023}. Note that despite of the primary role of the $d_{z^2}$ pairing, the induced $d_{x^2-y^2}$ pairing is equally important and the gaps on $\alpha$ and $\beta$ Fermi surfaces may also be large for sufficiently strong hybridization under high pressure.

The above mean-field equations yield a large $\Delta _{d}\sim J_{\perp }$ and a transition temperature $T_{\text{MF}}$ which is much higher than the
true $T_{c}$ due to the Cooper pair phase fluctuations in two dimension. On the other hand, 
the $d_{z^2}$ pairs alone have small superfluid stiffness ($\sim \delta_d t^d$) and may not give a large $T_c$ \cite{Lee-Nagaosa-Wen-2006}. A 
different estimate of $T_{c}$ may be obtained in the composite scenario by following the approach
outlined in Ref. \cite{Berg2008PRB} to apply a phase twist to the
kinetic energy term, $\epsilon _{\mathbf{k}}^{c}\rightarrow \epsilon _{%
\mathbf{k+q/2}}^{c}$, and calculate the phase stiffness $\rho _{s}=\partial
^{2}f/\partial q_{x}^{2}$, where $f$ is the free energy density and $q_{x}$
is the applied phase twist along $x$ direction. Without going into more
details, a qualitative analysis yields \cite{Berg2008PRB}
\begin{equation}
\rho _{s}(T)\sim \frac{\tilde{\Delta}_{c}^{2}}{T^{2}}t^{c}\sim \frac{\delta_d^2V^{4}}{%
J_{\perp }^{2}T^{2}}t^{c},
\end{equation}
which is valid for $g_tV^{2}/J_{\perp }\ll T\ll J_{\perp }$. Using the
condition, $\rho _{s}(T_{c})=\frac{2}{\pi }T_{c}$, for the two-dimensional
Kosterlitz-Thouless transition \cite{Kosterlitz-Nelson}, one gets eventually
\begin{equation}
T_{c}\sim t^{c}\left(\frac{\delta_dV^2}{J_{\perp }t^{c}}\right)^{2/3},
\label{Tc}
\end{equation}
which supports a high $T_c$ for finite $\delta_d$ as in La$_3$Ni$_2$O$_7$ under high pressure and explains the absence of superconductivity at ambient pressure because of the full occupancy of the $d_{z^2}$ bonding band ($\delta_d=0$). Our theory also predicts that $T_c$ may be enhanced in the low pressure region of the superconductivity by introducing more holes on the $d_{z^2}$ bonding band. On the other hand, applying higher pressure can be regarded as enhancing the hybridization $V$. While a larger $V$ may promote the phase stiffness, it also tends to compete with the local pairing field, causing non-monotonic variation of $T_{c}$ \cite{Berg2008PRB}. This is confirmed in Monte Carlo simulations of the minimal model \cite{Qin2023}, in good accordance with experimental observations 
\cite{Sun2023,Yuan2023,Cheng2023}.

\begin{figure}[t]
\begin{center}
\includegraphics[width=0.4\textwidth]{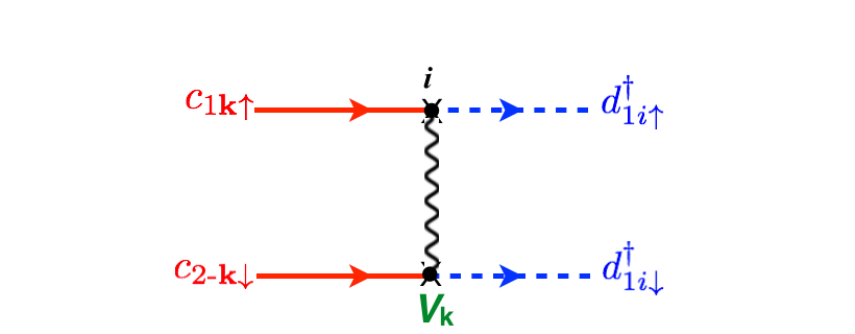}
\end{center}
\caption{Illustration of the composite pairing mechanism with strongly coupled $d_{z^2}$ and $d_{x^2-y^2}$ inter-layer spin singlet pairs due to hybridization, where $d_{z^2}$ pairs are more local and have small phase stiffness. $V_\mathbf{k}$ denotes their hybridization.}
\label{fig3}
\end{figure}

Our two-component theory represents the lowest-order contribution to the pairing in the presence of self-doped holes in the $d_{z^2}$ orbitals, as illustrated in Fig. \ref{fig3}, with the local inter-layer $d_{z^2}$ spin singlets playing the primary role and attaining phase coherence through hybridization with the metallic $d_{x^2-y^2}$ bands. This mechanism is robust against other perturbations such as inter-layer hybridizations or intra-layer spin interactions. A Hund scenario has been proposed previously in some other superconducting systems  \cite{Georges2013ARCMP,Han2004PRB,Dai2008PRL,Puetter2012EPL,Csire2018EPJB,Kostin2018NM,Ghosh2020PRB,Roig2022PRB}, but we believe it cannot be the primary driving force for the electron pairing in La$_3$Ni$_2$O$_7$ since no superconductivity is observed at ambient pressure with fully occupied $d_{z^2}$ bonding band. For the similar reason, we also do not think that spin fluctuations may play an important role. More systematic numerical and experimental studies are needed to settle all the debates.

\textbf{Discussion and conclusions.}-- While our theory is motivated by the
composite scenario for high-$T_c$ cuprates \cite{Kivelson2002},
the two systems have some important differences. First, La$_3$Ni$_2$O$_7$ is
a multi-orbital bilayer system. This provides a natural basis for the two-component
theory, where the more localized $d_{z^2}$ electrons act as the pairing
component and the itinerant $d_{x^2-y^2}$ electrons provide the metallic
component. While in cuprate high-$T_c$ superconductors, only one orbital is involved (at least for the Hubbard model), causing some complications in theory. Second, in contrast to the cuprate superconductivity
whose parent state is an antiferromagnetic Mott insulator, the high-$T_c$
superconductor La$_3$Ni$_2$O$_7$ exhibits no long-range antiferromagnetic
order. Rather, it starts with well-formed local spin singlets of Ni-$d_{z^2}$
electrons and superconductivity is induced by doping into the spin-singlet array. This has an obvious advantage, because antiferromagnetic ordering typically competes with the superconductivity, causing failures in
previous attempts to find more high-$T_c$ superconductors guided by the
cuprate-Mott mechanism.

To summarize, we have proposed a strong coupling picture for La$_{3}$Ni$_{2}$%
O$_{7}$ under high pressure and a two-component pairing mechanism for its high-$%
T_{c}$ superconductivity. We show that the weak coupling picture based on
almost fully-filled Ni-$d_{z^{2}}$ bonding band may be incorrect due to the
strong on-site Coulomb interaction, and construct a minimal effective model
for describing its low-energy physics. This leads to the idea of local
inter-layer spin singlet pairing of Ni-$d_{z^{2}}$ electrons due to a large
inter-layer superexchange interaction via the apical O-$p_{z}$ orbital. High-%
$T_{c}$ superconductivity may then be established by hybridization with the
metallic $d_{x^{2}-y^{2}}$ electrons to induce the global phase coherence.
Our theory highlights the importance of the bilayer structure of superconducting La$_{3}$Ni$_{2}$%
O$_{7}$ and provides
an alternative route to explore more high-$T_{c}$ superconductors \cite{Gao-Lu-Xiang2015}. Our model
provides a basis for more elaborate theoretical investigations.

\acknowledgments
One of the authors (GMZ) acknowledges stimulating discussions with X. J.
Zhou on trilayer cuprates. This work was supported
by the Strategic Priority Research Program of the Chinese Academy of
Sciences (Grant No. XDB33010100), China's Ministry of Science and Technology (Grant No. 2022YFA1403900), and the National Natural Science Foundation of China (Grant  No. 11920101005).

\textbf{Note added.} When preparing this manuscript, we notice several
preprints \cite{Congjun2023,YaHuiZhang,Yurong2023,Weili2023} which studied a 
bilayer $t$-$J$ model with strong inter-layer coupling and considered the 
possible superconducting pairing symmetry. Several later preprints \cite{Jiang2023,Tian2023,Lu2023,Zhang2023} also emphasized the importance of $d_{x^2-y^2}$ pairing for the superconductivity. F.C.Z. was partially supported by the Chinese Academy of Sciences under contract No. JZHKYPT-2021-08.

\end{document}